# MEASUREMENTS OF RF CAVITY VOLTAGES BY X-RAY SPECTRUM MEASUREMENTS

J P Duke, A P Letchford and D J S Findlay,

Rutherford Appleton Laboratory, Oxfordshire OX11 0QX, England

*Abstract*

Measurement of the endpoints of X-ray spectra emitted from RF cavities is a useful non-invasive technique for measuring peak voltages within the cavities. The matching of calculated X-ray spectra to measured spectra is described, with emphasis upon the endpoint region.

## 1 INTRODUCTION

The peak voltage reached within an RF cavity is one of the most important parameters of the cavity. But such voltages can be difficult to measure, since the introduction of probes generally changes electromagnetic field distributions and alters the voltage being measured. However, X-rays generated by the stopping in metal electrodes of stray electrons accelerated across voltage gaps are unaffected by local electromagnetic fields, and can easily escape through the cavity walls to be detected outside. X-rays can therefore be used for non-invasive measurements of voltages within RF cavities.

Excluding production of characteristic X-rays, an energy spectrum of X-ray photons decreases monotonically as photon energies increase, and goes to zero at the energy of the electrons generating the X-rays. A measurement of the electron energy is therefore made by measuring the X-ray spectrum and locating its endpoint energy. But locating the endpoint energy of an experimentally measured spectrum can be difficult because of statistical uncertainties in the data and the presence of background. This paper presents a method of calculating the expected X-ray spectrum shape which can be fitted to data to help locate endpoints more reliably.

## 2 X-RAY SPECTRA CALCULATIONS

Ideally, calculation of an X-ray spectrum from an RF cavity would take into account the angular dependence of the X-ray photons relative to their generating electrons, but because it adds great complexity to the calculation, and because anyway it is often not clear what the distribution of angles of the electrons themselves are, the present calculation is restricted to integrated-over-angle spectra. By considering the material in which an electron slows down and stops as a series of slices of thickness $dx$, the integrated-over-angle X-ray spectrum as a function of photon energy $k$ for a fixed electron energy $E$ can be written as $dY/dk(E,k) = (N_A/A) \int d\sigma/dk(E'(x),k)\, dx$ where $dY/dk$ is the number of X-ray photons per unit energy interval per electron, $N_A$ is Avogadro's number, $Z$ and $A$ are the atomic weight and number of the material, $d\sigma/dk$ is the integrated-over-angle bremsstrahlung cross-section, $E'(x)$ is the electron energy after the electron has travelled a distance $x$ (g/cm²) in the material, and the integral extends from $0$ to $x_k$ where $x_k$ is the thickness at which the electron energy has reduced from $E$ to $k$. The bremsstrahlung cross-section $d\sigma/dk$ has been tabulated by Pratt *et al.* in [1], and the tabulated numbers may be represented by $(\beta^2/Z^2)\, k\, d\sigma/dk = a\,(1 - b(k/E))$ where $a \approx 10$ millibarns ($10^{-27}$ cm²), $b \approx 0.5$, and $\beta$ is the velocity of the electron as a fraction of the velocity of light. The electron stopping power [2] in the material may be written $dE/dx = cE^d$, whereupon $E'(x) = (E^{(1-d)} - (1-d)cx)^{(1/(1-d))}$.

Since in an RF cavity the voltages are oscillating sinusoidally, the X-ray spectrum above must be integrated over a sinusoidally varying electron energy, giving $dY'/dk(E_0,k) = (2/\pi) \int dY/dk(E_0 \sin\theta, k)\, d\theta$ where the integral extends from $\sin^{-1}(k/E_0)$ to $\pi/2$, and $E_0$ is the peak value of the oscillating voltage.

Fig. 1 compares an X-ray spectrum calculated as described above for a peak voltage of 90 kV between copper electrodes and the very simple, convenient but inevitably approximate X-ray spectrum of Kramers [3] $dY/dk = ((2 \times 5 \times 10^{-4}\, Z) / 511)\,(E - k) / k$ (photons per keV). It can be seen that while the two spectra are similar at low energies, they increasingly diverge, by up to an order of magnitude, as the photon energy increases.

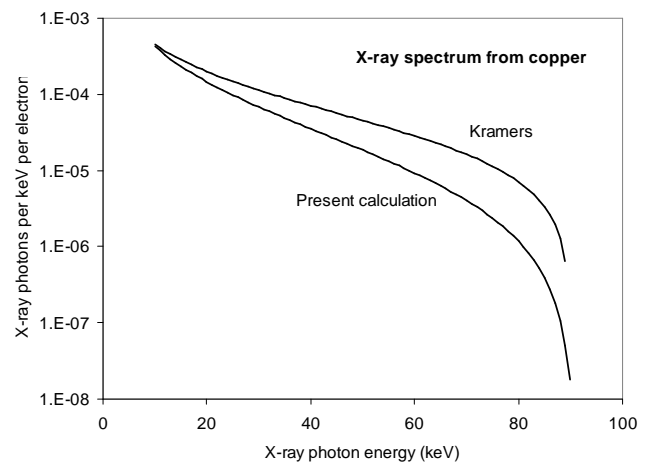

Figure 1  Comparison of X-ray spectra calculations

# 3 COMPARISON WITH MEASURED X-RAY SPECTRA

Comparison of the above calculations has been made with X-ray spectra measured from a 202.5 MHz 4-rod RFQ [4,5] built by the Institut für Angewandte Physik at the Johann Wolfgang Goethe - Universität in Frankfurt as part of a collaboration between the university and the Rutherford Appleton Laboratory (RAL). A preliminary version of the above calculations has already been used in [6]. The design peak rod-to-rod voltage in the RFQ is 90 kV for an input power of 200 kW.

The X-ray spectrum from the RFQ was measured using a 20% EG&G/Ortec GMX high purity germanium (HPGe) photon spectrometer with a Schlumberger Enertec PSC 761R preamplifier and a Canberra 2020 spectroscopy pulse shaping amplifier, the latter being connected to an EG&G/Ortec TRUMP-8K-32 pulse height analysis (PHA) card TRUMP in a PC. A 1.6 mm diameter hole was drilled in the 5 mm lead shield completely surrounding the RFQ to allow X-ray photons to fall on the detector. The calibration of the system below ~100 keV was established using the 26.3 and 59.5 keV lines from a small $^{241}$Am radioisotope source. The RFQ was driven at 50 pps with RF pulse lengths of 300–400 µs. The X-ray photon count rate in the detector was ~3 counts per 300–400 µs pulse.

Fig. 2 shows the X-ray spectrum from the RFQ at a power of 150 kW measured over 30 minutes. The HPGe detector was not shielded, nor was the PHA card gated around the RF pulse, so a background run with the small hole through the lead shield closed was taken and subtracted from the data measured with the hole open.

To compare with these data, an X-ray spectrum shape was calculated as described above. The RFQ rods are copper, so Z = 29 and A = 63.5. The parameters a and b were taken to be 9 mb and 0.5 respectively. The parameters c and d were taken to be $5.85\times10^4$ (in units consistent with E being in keV) and –0.664 respectively, which when used in $dE/dx = cE^d$ represent the electron stopping power in copper to within 8% everywhere between 10 and 200 keV. To account for X-ray absorption in the 4 mm thick RFQ vacuum vessel, the spectrum was then multiplied by $\exp(-(\mu/\rho(k))\,x)$ where $\mu/\rho$ is the mass attenuation coefficient (cm²/g) for photon absorption for stainless steel (essentially iron) obtained from [7]. To account for the finite resolution of the HPGe detector, a gaussian of 3 keV FWHM was folded into the spectrum. To account for the small differences amongst the four pairs of inter-rod voltages, four quarter-strength spectra were summed with peak voltages $f_{12}V_{nom}$, $f_{14}V_{nom}$, $f_{32}V_{nom}$ and $f_{34}V_{nom}$, where $f_{ij}$ is the factor by which the voltage between rods i and j averaged along the length of the rods is different from the nominal voltage $V_{nom}$; values of $f_{14}$, $f_{14}$, $f_{32}$ and $f_{34}$ were 1.00, 1.07, 0.93 and 1.00 respectively. The resultant spectrum shape was then fitted to the data, allowing the nominal peak voltage or endpoint energy to float. It can be seen from Fig. 2 that the fit is reasonably good, and a value of 69.2±0.5 keV was obtained for $V_{nom}$.

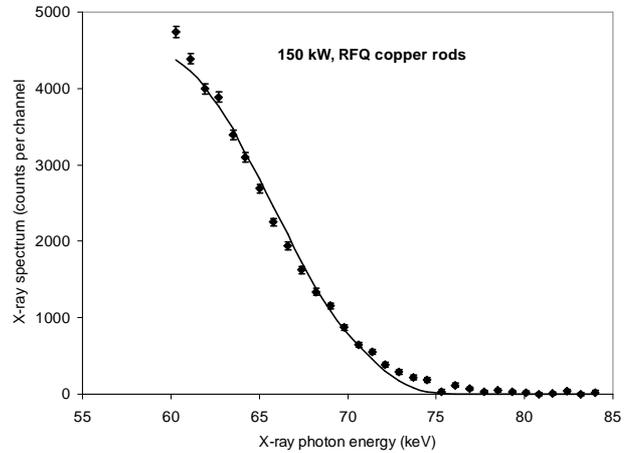

Figure 2  Comparison of measured and calculated X-ray spectra

It is evident from Fig. 2 that there are small discrepancies between the shapes of the measured and calculated spectra. While these discrepancies may be reduced when the spectrum is re-measured with a gated PHA card, or when the RFQ is fully conditioned, it is more likely that the discrepancies are simply due to approximations made in the calculations. Nevertheless, by fitting spectrum shapes calculated as described above to measured spectra, is evident that peak voltages or endpoint energies may be extracted to within an uncertainty of ~1 keV.

## REFERENCES

[1] R H Pratt *et al.*, Atomic Data Nucl. Data Tables **20** (1977) 175
[2] Studies in Penetration of Charged Particles in Matter, National Academy of Sciences - National Research Council, Washington, DC, Nuclear Science Series Report no. 39, NAS-NRC 1133 (1964)
[3] H W Koch and J W Motz, Rev. Mod. Phys. **31** (1959) 920
[4] H Vormann *et al.*, "A New High Duty Factor RFQ Injector for ISIS", EPAC 98, page 782
[5] A P Letchford and A Schempp, "A Comparison of 4-Rod and 4-Vane RFQ Fields", EPAC 98, page 1204
[6] A P Letchford *et al.*, "First Results from the ISIS RFQ Test Stand", paper THP4A05, EPAC 2000
[7] J H Hubbell, Photon Cross Sections, Attenuation Coefficients and Energy Absorption Coefficients From 10 keV to 100 GeV, NSRDS-NBS 29 (1969)